\documentclass[preprint,12pt]{aastex}
\usepackage{epsfig}
\usepackage[dvips]{color}
\def\pr{progenitor}

\shorttitle{Spin up/Spin down type Ia supernovae}
\shortauthors{Di Stefano et al.}

\begin{document}


\title{Spin-Up/Spin-Down models for Type Ia Supernovae}

\author{R. Di\thinspace~Stefano$^1$, R. Voss$^2$, J.S.W. Claeys$^3$}
\affil{$^1$Harvard-Smithsonian Center for Astrophysics, 60
Garden Street, Cambridge, MA 02138, $^2$Department of Astrophysics/IMAPP, 
Radboud University Nijmegen, PO Box 9010, NL-6500 GL Nijmegen, the 
Netherlands, $^3$Sterrekundig Instituut, Universiteit Utrecht, 
PO Box 800000, 3508 TA Utrecht, The Netherlands}

%%%%%%%%%%%%%%%%%%%%%%%%%%%%%%%%%%%%%%%%%%%%%%%%%%%%%%%%%%%%%%%%%%%%%%%%%%%
%
%                   ABSTRACT
%
%%%%%%%%%%%%%%%%%%%%%%%%%%%%%%%%%%%%%%%%%%%%%%%%%%%%%%%%%%%%%%%%%%%%%%%%%%%

\begin{abstract}
In the single degenerate scenario for Type Ia supernova (SNeIa),
a white dwarf (WD) must gain a significant amount of matter from a
companion star. Because the accreted mass carries angular momentum,
the WD is likely to achieve fast spin periods, which
can increase the critical mass, $M_{crit}$, needed for explosion. When 
$M_{crit}$ is higher than the maximum mass achieved by the WD, 
the WD must spin down before it can explode. This introduces a delay 
between the time at which the WD has completed its epoch of mass gain  
and the time of the explosion. Matter ejected from the binary during mass transfer therefore has a chance to 
become diffuse, and the explosion occurs in a medium with a density similar to
that of typical regions of the interstellar medium. 
Also, either by the end of the WD's mass increase or else by the time of 
explosion, the donor may exhaust its stellar envelope and become a WD. 
This alters, generally diminishing, explosion signatures related to the donor star.
Nevertheless, the spin-up/spin-down model is highly predictive.
Prior to explosion, progenitors can be 
super-$M_{Ch}$ WDs
in either wide binaries with WD companions, or else in
cataclysmic variables.  
These systems can be discovered and studied  
through wide-field surveys. Post explosion, the spin-up/spin-down model  
predicts a population of fast-moving WDs, low-mass stars, and even
brown dwarfs. In addition, the spin-up/spin-down model provides a 
paradigm which may be able to explain both the similarities and 
the diversity observed among SNeIa.
\end{abstract}

\maketitle

\section{Introduction}

Type Ia supernovae (SNeIa) are believed to be the explosions of 
carbon-oxygen white dwarfs (CO WDs). To explode, a CO WD must reach 
a critical mass ($M_{crit}$) generally assumed to be 
the Chandrasekhar mass ($M_{ch}\sim1.4$ $M_{\odot}$). This can
be achieved either through accretion from a companion star (the
single-degenerate [SD] scenario) or through the merger of two WDs
(the double-degenerate [DD] scenario).
Key signatures of the SD scenario 
include direct detection of progenitors in
archival images, direct detection of companions in
supernova remnants, and radiation emitted when light and matter
from the supernova interact with the companion star or with 
circumstellar material that
had been ejected from the progenitor binary. With the exception of 
signatures that may be due to absorption by circumstellar
material in a small number of SNeIa \cite[]{Patat2007}, 
these strong signatures have not yet been definitely 
detected, calling into question the relevance of SD models.  

In SD models, the WD must accrete and 
retain matter. 
This requires high mass infall rates, with  
$\dot M > 10^{-7} M_\odot {\rm yr}^{-1}$ \cite[] {Iben1982, Nomoto1982, 
Prialnik1995, 
 Shen2007}.  
Because infalling matter carries angular momentum, the angular momentum
of the WD must increase. 
Although spin-up seems certain to occur,
its effects are difficult to compute from first principles. 
One effect is an increase in the 
value of $M_{crit}$ \citep{Anand1965,
Roxburgh1965,Ostriker1968,Hachisu1986,Yoon2005}. 
We will show that an increase in $M_{crit}$ 
has a profound effect on the 
progenitor signatures. 
Some of the oft-expected donor signatures are diminished, 
possibly explaining why they have either not been detected, or have been
detected only rarely.
Nevertheless, the spin-up/spin-down model is testable, because it 
suggests alternative ways to identify the progenitors and test SD models. 
In section \ref{sect:model} we discuss
the model, using 4 key points to summarize
the features relevant to observations of the progenitors and explosions, to which we turn  
in section \ref{sect:effects}. 
In section \ref{sect:paradigm} we discuss how spin-up/spin-down 
provides a testable paradigm that can explain 
both the unity and diversity among
SNeIa. 
 
\section{Spin-Up and Spin-Down}
\label{sect:model}
%\textcolor{red}{starting sentence: see Rasmus version}
\noindent(1) \textit{Infalling matter spins up the WD to near-critical 
rotation.} 
Because infalling matter carries angular momentum, the angular momentum
of the WD must increase when the infalling matter is retained.  
Spin-up is a common process in accreting compact objects. 
Neutron stars (NSs), for example, can
be spun up to periods of a few milliseconds 
\citep[see review by][]{Lorimer2008}. Similarly a number
of fast-spinning WDs that must have been spun up by accretion are known, 
for example WZ~Sge, 27.87~s; AE~Aqr, 33.06~s; V842~Cen, 56.82~s and 
V455~And, 67.2 s \footnote{see the catalogue of intermediate polars at
http://asd.gsfc.nasa.gov/Koji.Mukai/iphome/iphome.html}. These periods are much longer than for NSs,
due to the much higher moment of inertia of WDs, but similar to the
milli-second pulsars, the surface velocity is only a factor of few
lower than the escape speed.

We can measure the spins in these specific systems because
the WDs are intermediate 
polars (IPs) where the accretion is channeled along the field lines of the WD
\citep{Warner1995}.
This is possible only for relatively modest accretion rates; higher
rates will increase the infalling matter density and probably quench 
the magnetic fields. 
The binaries most likely to be progenitors of SNeIa  have 
rates of mass transfer that are hundreds or thousands of times greater
than those inferred for IPs. 
%\footnote{Two supersoft x-ray sources in M31 exhibits 
%periods of 217.7 and 865.5~s \citep{Osborne2001,Trudolyubov2008}. It is possible that the period is a spin period and 
%that the source is a nuclear-burning WD. It is not presently possible to 
%verify these conjectures.}
The retention of mass should 
make it possible to spin mass-gaining WDs to even shorter periods than 
measured for IPs, even if measurements are difficult.  

GK~Per, which experienced 
a classical nova in 1901, has a spin period of 351~s,
and is spinning up at a rate measured to be  $0.00027\pm.00005$s~yr$^{-1}$
\citep{Mauche2004},
corresponding to a spin-up of $2.7 \times 10^5$~s per solar mass accreted in this system.  
The WDs which evolve towards SNeIa must accrete {\sl at least} 
$0.2\, M_\odot.$ Although the specific angular
momentum carried by infalling matter will vary among binary systems
[see, e.g., \citet{Popham1991}],
the spin-up of GK~Per suggests that WDs can gain 
enough angular momentum to reach critical rotation.

\noindent(2) \textit{The rotation increases the critical mass $M_{crit}$, 
needed for the explosion.} 
This implies that accreting WDs can achieve masses in excess of $M_{Ch}$ 
without either exploding or imploding. 
For a rigid rotator, maximal rotation produces an increase in $M_{crit}$ 
of roughly $5\%$ \citep{Anand1965,Roxburgh1965}. For more complex 
radial distributions of the internal angular momentum, \citet{Ostriker1968}
showed that the critical mass could become
very high; they constructed models with $M_{crit}$ as high as $4\, M_\odot,$
noting, however, that not  all of the configurations they considered
were likely to be realized in nature. 
\citet{Hachisu1986} also found stable equilibrium configurations with 
$M_{WD} > 2\, M_\odot.$ \citet{Yoon2005} considered
spin-up due to accretion and derived comparably high masses.
\citet{Piro2008} included viscous effects and found that, 
under certain input assumptions, the WD should be able to achieve a state
close to uniform rotation  
during much of the accretion phase, 
but that differential rotation could be important
during a short-lived ($\sim 10^3$ years) ``simmering'' phase just 
prior to explosion. The bottom line is that the values of $M_{crit}$
are difficult to compute from first principles, and that the rigid-rotation
limit can be taken to give a lower bound.  
%Although these 
%calculations were not able to take into account 
%all relevant physical effects,  
%including magnetic fields, they clearly indicate that maximal
%rotation can significantly increase the 
%value of $M_{crit}$. Since it is difficult to compute $M_{crit},$ the
%observational tests we will discuss in \S 3 could provide a way to
%determine the range of values possible for $M_{crit}$ in real situations.  

\noindent(3) \textit{Spin-down can occur when  $\dot M$ is low or 
when mass transfer has ceased.}  
As  $\dot M$ decreases, 
more angular momentum may be lost per unit time than gained
\citep[as is seen in AE Aqr][]{Meintjes2002,Ikhsanov2004}. 
 Gravitational
radiation associated with spin-induced
effects can produce spin-down in even isolated WDs.  
\citep{Sedrakian2006}.
Spin-down times are uncertain, but almost certainly exhibit a large range,
from $<10^{6}$ years to $>10^{9}$ years
\citep{Lindblom1999,Yoon2005}.

\noindent (4) \textit{Explosion occurs when the spin period has been 
reduced to a critical value, $P_{crit}.$}
As the spin of the super-Chandrasekhar WD decreases, so does the value
of $M_{crit}$. When the value of $M_{crit}$ falls below the current mass
of the WD, the WD will explode or, if it has crystallized,
it may collapse \cite[]{NomotoKondo1991}. The nature of the event 
and its
appearance depend on the state of the WD when mass gain stops and on the details of the subsequent evolution.
For example, does mass transfer continue at
a low level? what is the time required for spin down?   
%This requires that the WD has not yet
%crystallized (in which case the outcome is an accretion-induced collapse
%into a neutron star \cite[]{NomotoKondo1991}). However, the timescale for the WD to cool to such
%low temperatures is $geq 10^9$ years \citep{Yoon2005}. Furthermore, if accretion
%continues at a lower rate, crystallization can be avoided altogether.

\section{Observational Signatures}
\label{sect:effects}
\subsection{Background}

\noindent{\sl Population:} 
Let $f$ denote the fraction of all SNeIa \pr s in which (a)~spin up
produces a significant change in the value of  $M_{crit}$, and  
(b)~the maximum mass achieved
by the WD is smaller than  $M_{crit}$, necessitating  
an interval of spin down.
If the rate of SNeIa is $R,$
then the number of spinning-down progenitors in the Galaxy
is
$f\times (3\times 10^5) \, 
(\tau/10^8 {\rm yrs}) \times (R/0.003 {\rm yr}^{-1}),$
with up to a few thousand lying
within a kpc of Earth. $\tau$ is the spin-down time: the time between the
end of genuine mass gain by the WD and the explosion.  
There could be an even larger number of Galactic post-explosion systems:
$f \times (3 \times 10^7) \times (R/0.003 {\rm yr}^{-1})$, if SNeIa have been
occurring in the Galaxy for $10^{10}$ years.   

\smallskip

\noindent{\sl Binary Evolution:} 
SD SNeIa progenitors must have donor 
stars whose state of evolution, mass, and 
orbital separation enable them to contribute mass at high rates.
Giant donors can do this if the orbital separation is favorable. 
Once $\dot M$ from a giant donor is high enough to promote nuclear burning 
by the WD, it is likely to stay high. The binary will be a symbiotic in
which the WD can gain mass and 
angular momentum until the giant's envelope is depleted. 
The final pre-SNeIa state is a   
wide-orbit double WD.

The same evolutionary path can be followed when the donor starts 
mass transfer as a subgiant if its core is evolved enough. 
Less evolved subgiant donors and main-sequence (MS) donors follow
an alternative channel in which 
the mass 
ratio, $q$, between donor and WD plays an important role. 
The value of $\dot M$ can be high enough to promote nuclear burning 
only when $q>1$. When the mass ratio reverses, the 
rate of mass transfer decreases dramatically. 
The WD can begin to spin down. 
The donor may 
lose
a significant fraction of its remaining mass. 
Subgiant donors could become WDs, 
reproducing 
the signatures described above for giant donors.
For MS donors, the binary will become an accretion-powered
CV; 
long spin-down times would transform  
the donor into 
a degenerate object of brown-dwarf mass, 
with orbital period as low as $\sim 90$ minutes.
Figure 1 shows the evolution of a subgiant donor whose WD companion
gains enough mass to slightly exceed $M_{Ch}.$
  
\subsection{Progenitor signatures}

\noindent{\bf Missing Signatures:} 
Signatures thought to be integral parts of SD models are diminished. 
For example, even a spin-down time of $10^5$~years provides 
enough time for circumbinary material to dissipate. Furthermore, the
donors are likely to be either compact objects at the time of explosion or 
else low-mass stars. 
Signatures of interaction with the 
supernova would therefore tend to  
be diminished relative to the case in which spin doesn't play a role.
\footnote{Note that signatures related to 
circumbinary material and/or interactions with a companion can be
ambiguous. For example, a DD may take place inside
a common envelope if the envelope ejection efficiency is low. 
Or, if some SNeIa
(either SD or DD)
take place in high-order multiple systems, stars not directly involved in the
explosion may produce detectable signatures.}  
In addition, the donors tend to be dim, making them difficult to detect,
especially in external galaxies. Nor are the WDs likely to be burning
nuclear fuel just prior to explosion. This is consistent with the small numbers
of supersoft x-ray sources found in external galaxies  
\citep{RD2006, DiStefano2010a, DiStefano2010b, RD2010, 
RD_M31_2004, RD_M104_2003}.\footnote{Nuclear burning should take place, however,
while the WD gains mass. The lack of SSS-like emission, may be 
due to an extended
photosphere or to absorption by circumstellar matter that then dissipates prior
to explosion \citet{DiStefano2010a,DiStefano2010b}.    
}

\noindent{\bf Tests of the models:} Systematic searches of
data from wide-field surveys,
including SDSS, Pan-STARRS, and LSST, should be able  
to identify those Galactic progenitors nearest to us.
[See, e.g.,  \citet{Kleinman2007,Szkody2006}  for SDSS-based identification of
WDs, and CVs, respectively.]
To test spin-up/spin-down models, we want to 
measure the mass function of the 
spinning-down WDs. The maximum mass will
tell us the maximum value of $M_{crit},$ testing whether
differential rotation occurs.
Even should no super-$M_{Ch}$ WDs be found, the mass distribution
would provide hitherto unavailable information on the mass gain during
binary evolution.  

{\sl Wide double WDs:} 
Binaries containing a super-$M_{Ch}$ in a wide orbit with
a compact companion are  
distinctive, in that they exhibit the
spectra of two hot WDs (Figure 2). 
The lower the
mass of the secondary, the cooler it will be, and the larger will be the
spectral contrast. Studies which have identified WD/M-dwarf pairs in 
data from e.g.,  
SPY \citep{Maxted2007} demonstrate that it will be possible to either identify
or place limits on the existence of the wide double-WD progenitors
we predict. 
The double-WD SNeIa progenitors
 with the smallest spectral contrast would be those in which
the secondaries are the most massive. These would, however,
be distinctive in another way: the separation between the two components 
could be resolvable (the top panel of Figure 3). 
When the spectrum indicates that the secondary mass is also high,
follow-up observations to determine if the WDs can be resolved would also
be useful. 

{\sl CVs and other mass-transfer binaries:} 
Wide-field surveys, combined with x-ray-source catalogs, 
 can identify CVs and other mass transfer binaries.
It is interesting to note that AE~Aqr appears to have had an evolution that
mirrors what is expected for SNeIa \pr s. The key difference is that the 
WD's mass, while larger than typical of WDs,  is smaller than $M_{Ch}$
(see, e.g., Meintjes 2002).

\subsection {Detecting the Remnants of the Donors} 
The SNeIa releases the donor from orbit.
\citet{Hansen2003} and \citet{Justham2009}
considered cases in which the donor has generally not yet finished its
evolution at the time of explosion.  
In the spin-up/spin-down scenario, many donors will have lost their
envelopes prior to explosion; the binary will therefore be lighter 
and have a lower orbital
velocity (bottom panel of Figure 3). 
With the current observational sample 
\citep{Oppenheimer2001, Justham2009} it is not possible to 
verify that high-speed WDs and
isolated low-mass WDs are remnants of SNeIa explosions, or to
distinguish among models. New surveys, particularly those 
that allow high-proper-motion remnants to be identified, will provide more data.  
  
  A  unique feature of the spin-up/spin-down model is that, if the donor
started as a MS star and if  
$\tau$ is large, the donor will be a degenerate brown-dwarf-mass
object at the time of explosion. Its speed will be high: for a two-hour
period around a $1.6\, M_\odot$ WD, $v \sim 570$~km~s$^{-1}$. 
Although they constitute a small fraction of Galactic brown dwarfs (at most
 $10^{-4}-10^{-3}$), 
some of these objects could be discovered through their action as lenses,
if complementary data allow radiation from the brown dwarf to 
be detected \citet{DiStefano2008}  

\subsection{Other Connections with SNeIa observations and calculations}

SNeIa are used as cosmological probes. If explosions occurring at different 
cosmic times have different amounts of local absorption, this would introduce
a systematic uncertainty into measurements of the acceleration of the 
Universe.  
When the \pr\ is a super-$M_{Ch}$ WD that must
spin down before explosion, circumstellar material will play less of a role
regardless of redshift. If, therefore, $f$ is large, the systematic
uncertainty becomes less significant.  

It is important to note, however, that both SD and DD models predict that
SNe explosions can occur in a wide range of systems (footnote 2),
so that absorption could play a role in some. Even when spin up occurs,
if the mass gain outstrips the increase in $M_{crit}$,
the explosion could take
place during the epoch of high $\dot M.$

There is evidence that some SNeIa occur ``promptly'',  
within a few $\times 10^8$~years  after star formation,  [see e.g., 
\citet{Maoz2010}]. 
This places limits on the spin-down time for exploding WDs with high-mass 
donors.

\section{Spin-Up/Spin-Down: A new paradigm}
\label{sect:paradigm}

Conservation of angular momentum plays an important role 
in astrophysics. 
It allows NSs and black holes to be spun up to near maximal rotation. 
It seems almost certain 
that WDs can be similarly spun up. 
Indeed, given the variety of donors and accretion geometries exhibited 
in nature, spin-up can fail only if there is a fundamental physical principle 
that disallows it. As long as spin-up to near-maximal rotation occurs, some 
of the effects we discussed will occur. Although 
theoretical uncertainties make predictions difficult,  we have shown that
spin-up/spin-down has testable consequences.
The measurements we propose can therefore
 provide input for theoretical work. 
%The work that is needed includes realistic
%computations of $M_{crit}$ and calculations of the explosions 
%of rotating WDs.    

The spin-up/spin-down model
 appears capable of explaining 
the full
range of SNeIa properties. 
The mass, $M,$ of the WD at the time of explosion
is the first parameter that determines the observable characteristics.
Without spin up, SD explosions should occur soon after the
WDs reach a critical mass that is very close in value to $M_{Ch}.$
With spin-up, 
the value of $M$ is influenced by the properties of the initial binary.  
For a rigid rotator,
the WD masses should 
lie in the range $M_{Ch}-1.05\, M_{Ch}.$ In other models, the mass can be
higher. By identifying the maximum WD mass, we will learn about the
angular momentum profile of the pre-explosion WDs.
Of course, only  a small  fraction of  donors can
provide enough mass to allow the WD to significantly exceed $M_{Ch}$;
thus, the largest number of pre-explosion WDs should have masses very close to
$M_{Ch}.$ By measuring the distribution of primary WD masses, we will therefore
learn about the binaries whose evolutions produce SNeIa.

All other things being equal,
each value of $M$ would correspond to a specific value of $P_{crit},$
the spin at which the value of the critical mass would become
equal to $M$.  In fact, however, 
the angular momentum and internal states will differ 
at the time mass accretion halts, introducing
a difference in the values of $\tau$. Furthermore,
if there is residual low-level accretion, this also affects the spin down time.
Thus, the value of $P_{crit}$ may be viewed as a second parameter which 
influences the explosion characteristics.

Finally, the variety of conditions expected at the time when
high-$\dot M$ mass infall ceases, combined with a wide range of possible
spin-down evolutions can yield very different pre-explosion conditions.
These can in turn produce some truly unusual
light curve and spectral
evolutions.
While we cannot determine whether spin-up effects explain the characteristics
of any specific explosion, it is instructive to consider 
SN~2008ge (Foley et al.\, 2008), 
an SN~2002cx-type explosion, showing an
unusual light curve and pattern of spectral evolution. The chemical composition,
pre-explosion HST images, and lack  of star formation in the host galaxy
make it almost certain that SN~2008ge was the explosion of a WD.
Yet, the explosion
itself may have been different from most SNeIa.
A complete
deflagration or else incomplete burning have been invoked as
possible explanations.

Spin-up/spin-down produces a new paradigm for the progenitors
of SNeIa. Key elements can be tested through observations. 
While not all SNeIa progenitors may be SD, and not all SDs
may be significantly affected by spin up, it seems
inevitable that angular momentum plays a role in some of the progenitors.

\bigskip
\noindent {\bf Acknowledgements:} It is a pleasure to acknowledge 
useful comments from Ryan Chornock, Ryan Foley, 
Robert P. Kirshner, and Andrew MacFadyen.  
This work was supported in
part by NSF through AST-0908878 and by a research and development
 award from the Smithsonian 
Astrophysical Observatory. The authors would like to thank the Lorentz
Center for its hospitality during the workshop
``Observational signatures of type Ia supernova progenitors'', where this 
work was started.

\begin{figure*}
\begin{center}
\psfig{file=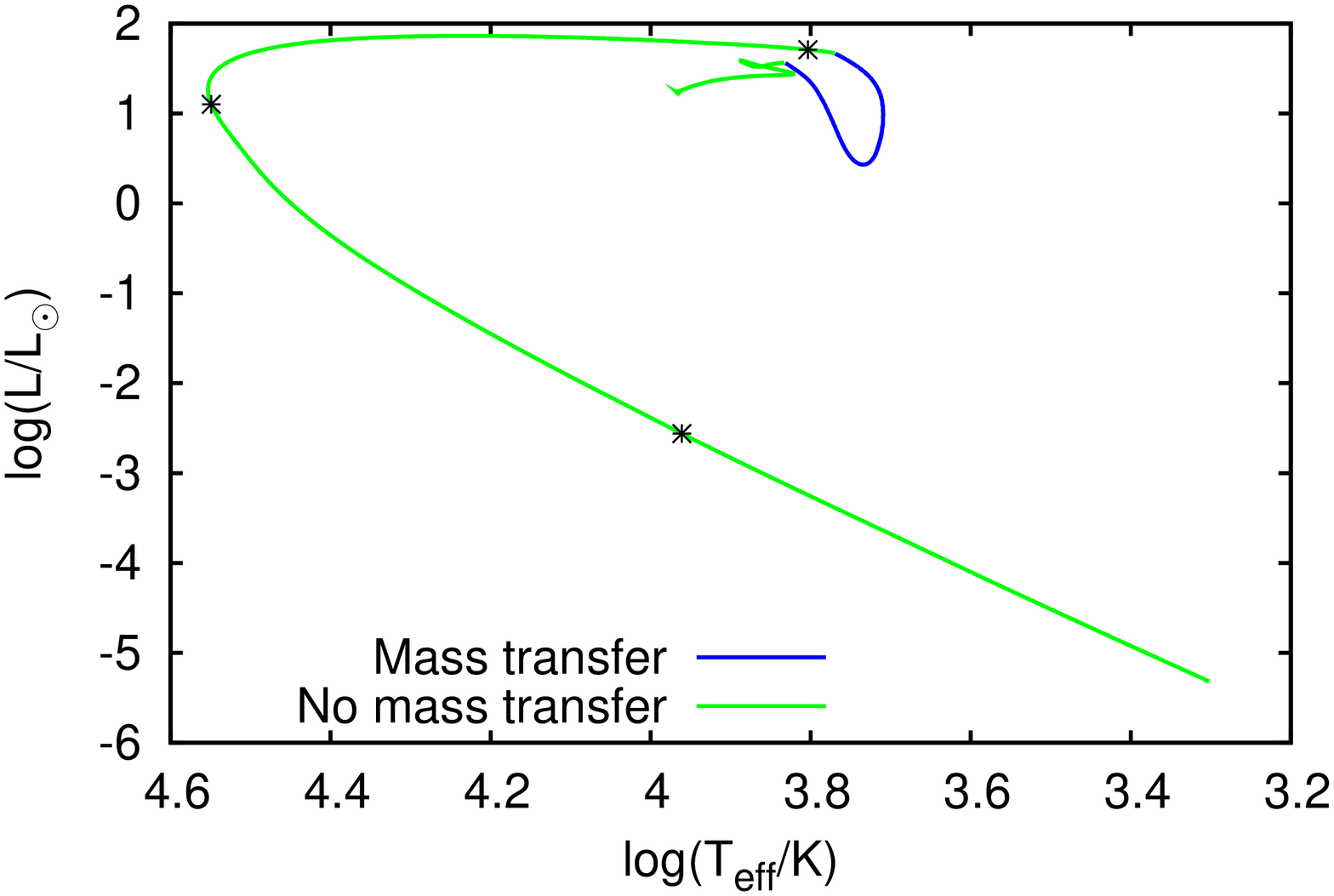,
height=6.0in,width=4.5in,angle=-90.0}
%\vspace{-2.1 true in}
\caption{
Hertzsprung-Russell diagram of the donor star in 
initial binary system of an $0.7\, M_\odot$ WD and and a $2\, M_\odot$ MS 
star (final: $M_{WD}  \sim  1.5\, M_\odot$, 
$M_{companion}  \sim  0.3\, M_\odot$). 
Red indicates the phase of mass transfer to the WD. 
Crosses indicate different times after mass transfer has 
ceased ($10^6, 10^7,10^9$ years).
Mass transfer starts when the donor is in the Hertzsprung gap and 
continuous during the GB. After mass transfer the 
donor star evolves into a He WD.
After $10^6$ years: the donor will appear as low-mass He-star. 
After $10^7$ years as a hot He-WD, after $10^9$ years as a cooler He-WD.
}
\end{center}
\end{figure*}

\begin{figure*}
\begin{center}
\psfig{file=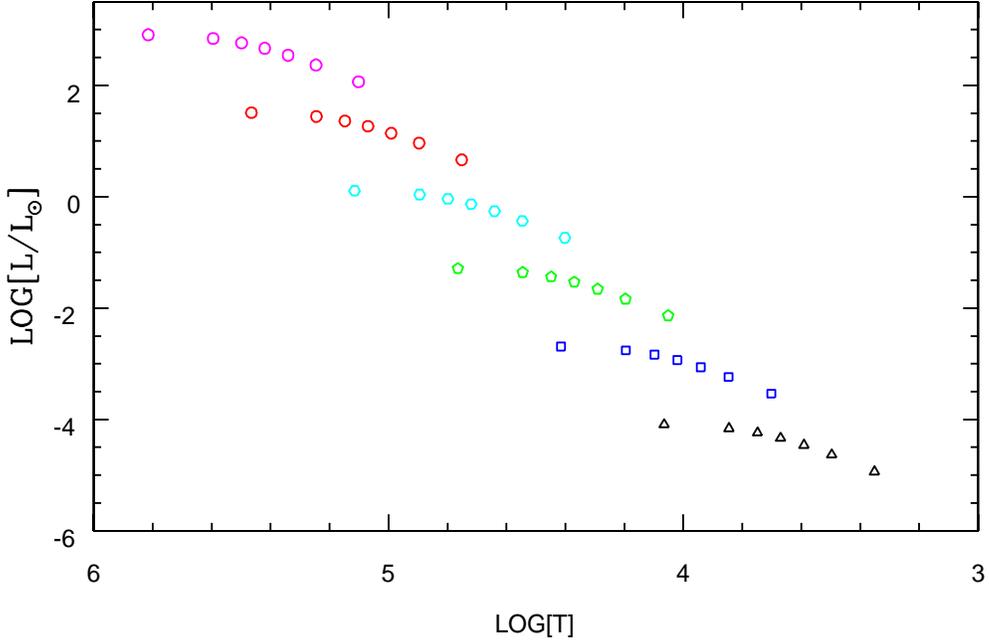,
height=7.5in,width=5.5in,angle=-0.0}
\vspace{-3.1 true in}
\caption{
Logarithm of luminosity versus the logarithm of temperature for
cooling
WDs in wide double-WD binaries. 
Each sequence of single-color points with a fixed number of 
sides corresponds to a given time after the end of mass transfer. 
The top row (magenta points) corresponds to $10^5$ years, and red, cyan, 
green blue, and black points correspond to 
$10^6, 10^7, 10^8, 10^9, 10^{10}$ years, respectively. 
The hottest 
and brightest system in each sequence corresponds to a Chandrasekhar-mass WD. 
The super-Chandrasekhar-mass primaries we consider may be somewhat hotter 
and brighter. Each subsequent point in the same-age sequence corresponds to a 
WD with a mass $0.2 M_\odot$ lower than the previous point of the sequence. 
The minimum mass shown is $0.2\, M_\odot.$ 
We used the realization of Mestel's cooling law
suggested by \citet{Kawaler1998}. Although this model does not 
exactly mirror the physical systems
we want to study, this figure illustrates  
the important feature that the massive WD is likely to be brighter 
and hotter than its lower-mass companion, and that both WDs are bright and 
hot compared with the majority of Galactic WDs, which are older. 
}
\end{center}
\end{figure*}
\begin{figure*}
\begin{center}
\psfig{file=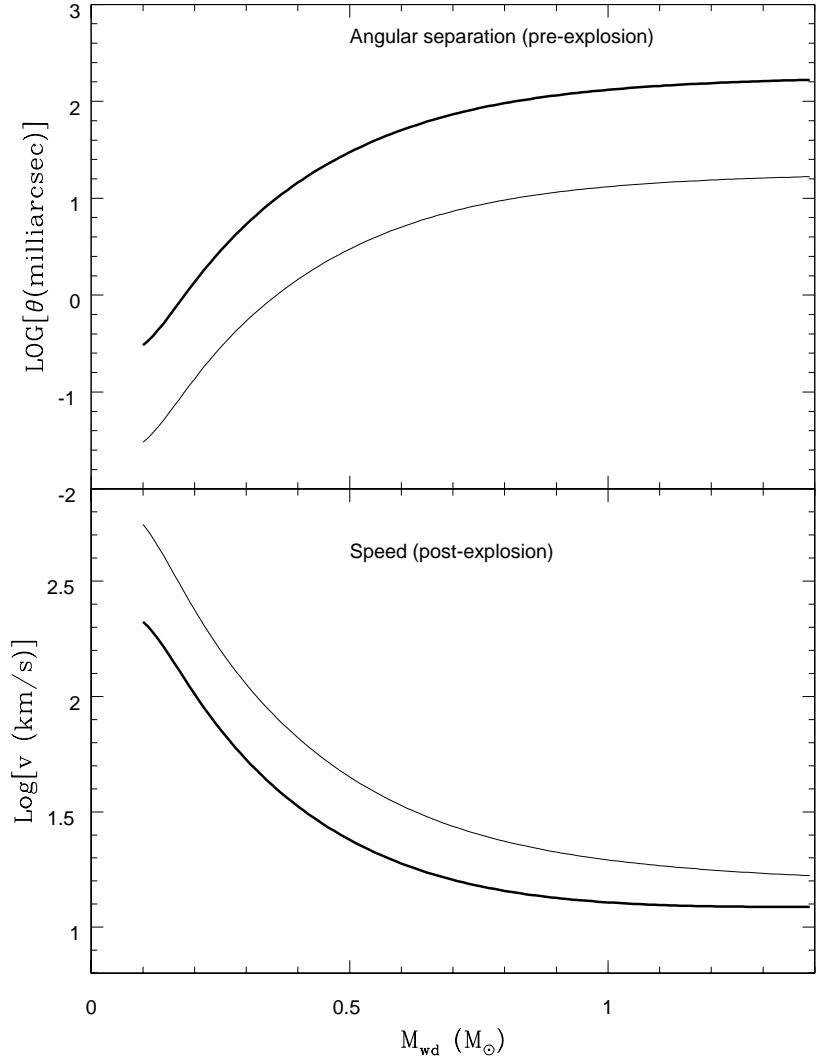,
height=6.0in,width=4.5in,angle=-0.0}
%\vspace{-2.1 true in}
\caption{
{\bf Top panel:} Logarithm of the angular separation (in mas) 
between the pre-explosion super-$M_{Ch}$ WD
and its less massive WD or pre-WD companion, as a function of the companion's 
mass. Top, dark curve: the distance to the binary is $100$~pc; bottom, lighter
curve: the distance to the binary is $1$~kpc. To compute the
orbital separation we assumed    
that the donor star 
is either a subgiant or giant that fills its 
Roche lobe until its envelope is exhausted. 
{\bf Bottom panel:} The logarithm of the 
speed of the companion is 
plotted as a function of the companion mass. Bottom, dark curve: results for the
spin-up/spin-down model, in which the donor has  lost its envelope prior
to the explosion.  
Plotted is the orbital speed (presumably close to the ejection speed) vs the core mass of the donor.
Upper, lighter curve is computed assuming that the donor
star has a total mass of $3\, M_\odot$ at the
time its WD companion explodes.} 

\end{center}
\end{figure*}

\end{document}